\documentclass[aps,prd,twocolumn,groupedaddress,showpacs]{revtex4}

\begin{document}


\title{The Cosmological Origin of Inertia: Mach's Principle}


\author{Christoph Schmid}
\email[]{chschmid@itp.phys.ethz.ch}

\affiliation{Institut f\"ur Theoretische Physik, ETH-H\"onggerberg, 
CH-8093 Z\"urich}

\date{\today}

\begin{abstract}
The axes of gyroscopes experimentally define local non-rotating frames,
i.e. the time-evolution of axes of inertial frames.
But what physical cause governs the time-evolution of gyroscope axes?
Starting from an unperturbed FRW cosmology with k = 0 
we consider cosmological vorticity perturbations (i.e. vector
perturbations) at the linear level, 
and we ask: Will cosmological rotational
perturbations exactly drag the axes of a gyroscopes relative to 
the directions of geodesics to galaxies in the asymptotic FRW space? 
Using Cartan's formalism with local orthonormal bases
we cast the laws of gravitomagnetism into a form showing the
close correspondence with the laws of ordinary magnetism. 
Our results,
valid for any equation of state for cosmological matter, are:
1) \nolinebreak The dragging of a gyroscope axis 
by rotational perturbations of matter
beyond the H-dot radius (H = Hubble constant) is exponentially suppressed.
2) \nolinebreak If the perturbation of matter is a 
homogeneous rotation inside a perturbation radius,
then exact dragging of the gyroscope axis by the rotational perturbation 
is reached exponentially fast 
as the perturbation radius gets larger than the H-dot radius.
3) \nolinebreak The time-evolution of a gyroscope axis 
exactly follows a specific average of the matter 
inside the H-dot radius.
In this sense Mach's Principle (that axes of local non-rotating
frames precisely follow some average of the motion of cosmic matter)
is a consequence of cosmology with Einstein Gravity.  
\end{abstract}


\maketitle


\section{Introduction}

{\it The observational fact.---} 
In tests of general relativity in the solar system,
two type of things are compared.
On the one hand
measurements of the precession of 
perihelia or of gyroscopes' spin axes
in Gravity Probe B
    \cite{GP.B}
{\it relative to distant stars and quasars.}
On the other hand the solutions of Einstein's equations 
for the solar system in 
asymptotic Minkowski space, 
which does not contain distant stars or quasars explicitely.
In this comparison the assumption is made (usually not spelled out)
that far outside the solar system,  
but much nearer to us than the nearest stars, 
the {\it local nonrotating frame} 
(of the asymptotic Minkowski space
in the solution of Einstein's equations)
is given by distant stars and quasars. 
This implicit assumption is tested to high accuracy
by the comparison of the observed perihelion precessions 
with the predictions of general relativity . 
This implicit assumption is a basic observational fact.
This observational fact has been called ``Mach0'' e.g. in
   \cite{Bondi},
although Newton already had written that a truly non-rotating system 
(established experimentally by Newton's bucket experiment) 
is approximately given by the stars 
   \cite{Barbour.Pfister}.
Since stars and quasars have proper motions, 
this formulation cannot be exact. 
But the proper motions of quasars
relative to the uniform Hubble flow 
are negligible for present-day tests of General Relativity.---
We conclude that the measurements of perihelion shifts  
and the measurement undertaken by Gravity Probe B  
are {\it tests of two things combined}, 
on the one hand tests of Einstein's equations in the solar system,
on the other hand tests of the principle ``Mach0''.


{\it Classical mechanics, special relativity, and general relativity  
for isolated systems in asymptotic Minkowski space} 
give no explanation of the
observational fact of ``Mach0'', 
except saying that this is an accident of initial conditions. 
In these theories one could have different initial conditions
where all stars could be in rotational motion
around us relative to our gyroscopes.---
Within these three theories the local non-rotating frames
(one aspect of inertial frames)
can be experimentally determined by axes of gyroscopes,
and conversely the time-evolution of gyroscope axes is
dictated by the laws of inertia, i.e. that the gyroscope
axes cannot rotate with respect to local inertial axes.
Hence in these three theories things are fully consistent 
but circular, and the question remains: 
What physical cause governs the time-evolution of 
the axes of gyroscopes and inertial frames?


{\it Mach's Principle.---} In the 1880's Mach
   \cite{Mach.Mechanik}
   \cite{Mach.Energy}
stated clearly and forcefully, as an alternative to 
Newtonian physics, the hypothesis that 
the axes of local non-rotating frames (i.e. axes of gyroscopes)
in their time-evolution 
{\it are determined by (are exactly dragged by, precisely follow)}  
``some average'' of the motion of matter in the universe.
This is what we take as the formulation of 
Mach's principle.---
Many alternative formulations of Mach's principle
have been proposed by other authors later.
We shall discuss Einstein's proposal
   \cite{Einstein.Mach} 
in a future paper.
Many of the alternatives to Mach's original formulation 
have been enumerated and briefly discussed by Bondi and Samuel 
   \cite{Bondi}.
Quite a number of these alternatives have almost nothing 
in common with Mach's ideas.


{\it Gravitomagnetism.---}
At the time of Mach there was no known mechanism,
by which matter in the universe could influence the motion of gyroscope axes.
With General Relativity 
came the needed mechanism, gravitomagnetism.
Thirring in 1918
   \cite{Thirring}
analyzed the partial dragging of the axes of inertial frames
inside a rotating infinitely thin spherical shell 
with uniform surface mass density and total mass $M.$ 
In the weak field approximation,
$G_{N}(M/R)_{\rm shell} \ll 1,$ he found that inside the shell 
the axes of local inertial systems at all points rotate 
relative to asymptotic Minkowski space with the same 
precession rate 
$\tilde{\Omega} = f_{\rm drag} \Omega_{\rm shell}.$ 
For the dragging fraction $f_{drag}$ he obtained 
$f_{\rm drag}= \frac{4}{3} G_{N}(M/R)_{\rm shell} \ll 1.$
There is only a tiny dragging effect,
unless $(M/R)_{\rm shell}$ approaches the value for a black hole.
To be relevant for Mach's hypothesis, 
{\it exact dragging} by the masses in the universe, 
{\it not a small influence,}
one must go to cosmological models.


{\it In this paper}
we analyze realistic cosmological models 
(as opposed to toy models) 
with realistic cosmological matter 
(as opposed to the contrived energy-momentum tensors 
discussed in the literature).
We start from an unperturbed Friedmann-Robertson-Walker 
(FRW) cosmology with k = 0, 
we add the most general cosmological vorticity perturbations 
(i.e. vector perturbations) at the linear level, 
and we ask: Will cosmological rotational
perturbations {\it exactly drag} the axes of any gyroscope 
relative to the directions of geodesics 
from the gyroscope to galaxies in the asymptotic FRW space?
In our analysis we cast the laws of gravitomagnetism 
into a form showing clearly the
close correspondence with the laws of ordinary magnetism. 
This is achieved by using Cartan's formalism with 
local orthonormal bases (LONBs) and fiducial observers (FIDOs).
Our results, stated in the abstract and presented in sections IV and V,
show that Mach's Principle (that axes of local non-rotating
frames {\it precisely follow} some average of the motion of cosmic matter)
is a consequence of cosmology with Einstein Gravity.
The crucial equation is  
Einstein's $G_{\hat{0} \hat{i}}$-equation, 
Amp\`ere's law for gravitomagnetism, 
     Eq.~(\ref{Ampere.FRW}) 
in a spatially flat FRW universe.
The mathematical statement, what average of the energy flow 
out there in the universe determines the time-evolution of gyroscopes'
axes here, is given in
     Eq.~(\ref{Mach}).
A short version of this work appeared in 
   \cite{my.gr-qc}.
In a subsequent paper we shall present the analysis for
vorticity perturbations on a FRW background with $k = \pm 1.$---
We use the conventions of Misner, Thorne, and Wheeler
   \cite{MTW}.


\section{Vorticity Perturbations, \protect\\    
         Fiducial Observers,      \protect\\    
         and the Gravitomagnetic Field}        

   {\it Cosmological Vorticity Perturbations.---} 
Linear cosmological perturbations, see J. Bardeen  
   \cite{Bardeen},
decouple in three sectors,
3-scalars (density perturbations), 
3-vectors (vorticity perturbations), and 
3-tensors (gravitational waves).
In the vector sector all quantities must be constructed from
a 3-vector field with vanishing divergence.
Hence the 3-scalar $\delta g_{00}$ is zero, 
the lapse function is unperturbed,
and the slicing of space-time in slices $\Sigma_{t}$
is unique, 
i.e. there is no gauge ambiguity about the time coordinate.
For linear vector perturbations
the intrinsic geometry of each 
$\Sigma_{t}$ remains unperturbed.
This holds because the perturbations of the Ricci scalar, 
of the space-space components and the trace of $T_{\mu \nu},$
and of the Ricci tensor all must vanish, 
and for $d=3$ the Riemann tensor 
can be built from the Ricci tensor.
Our choices of spatial coordinates are Cartesian 
for $k=0$
resp spherical FRW coordinates for $k=0, \pm 1$.
The line element is
\begin{equation}
ds^{2}=-dt^{2}+(ah_{i})^{2}(dx^{i})^{2}+
              2(ah_{i})^{2} \beta^{i} dx^{i}dt,
\label{line.element} 
\end{equation}
where $\beta^{i}$ is the shift 3-vector.
Geodesics on $\Sigma_t$ are straight lines on our choice of chart.
We consider vector perturbations in an asymptotic FRW universe,
i.e. $\beta^{i} \rightarrow 0$ for $r \rightarrow \infty.$
Our coordinates are fixed to ``distant galaxies'', 
i.e. to galaxies in the asymptotic unperturbed FRW space, and
the basis vectors in the coordinate basis,
$\bar{e}_i(P) = \partial / \partial x^{i},$ 
point along geodesics in $\Sigma_{t}$ from $P$
to fixed ``distant galaxies''.


   {\it Fiducial Observers.---}
Our aim is to obtain the laws of linearized gravito
magnetism in a form analogous to electromagnetism 
in a $3+1$ formulation. 
What is the operational definition 
for $\vec{E}_{g}$ (gravitoelectric field) 
and $\vec{B}_{g}$ (gravitomagnetic field)?
According to the equivalence principle 
for a free-falling, non-rotating observer there are 
no gravitational forces at his position,
$\vec{E}_{g}=0,\  \vec{B}_{g}=0.$
It all depends on the choice of fiducial observers, FIDOs, 
with their local ortho-normal bases, 
LONBs, see Thorne et al \cite{Thorne}.
Hence we work in the formalism of E. Cartan 
   \cite{MTW}.
Our choice of FIDOs: The world lines of our FIDOs are at fixed
$x^{i}$ in our coordinates, which are fixed to distant galaxies
in the asymptotic FRW universe, and 
$\bar{e}_{\hat{0}}(P) = \bar{u}_{\rm FIDO}(P)$. 
Hats refer to LONBs, and bars designate space-time vectors.
We choose the spatial basis vectors of our FIDOs, 
$\bar{e}_{\hat{i}}(P)$,
fixed to directions of geodesics on $\Sigma_t$ from $P$ 
to distant galaxies in the
asymptotic FRW universe. 
Specifically we fix $\bar{e}_{\hat{i}}(P)$ 
in the same spatial directions as 
$\bar{e}_{i}(P) \equiv \partial/\partial x^{i},$
i.e. in 4-space the directions of $\bar{e}_{\hat{i}}$ and
$\bar{e}_{i}$ differ by a pure Lorentz boost, see 
   Eq.~(\ref{expansion of basis vectors}) below.
The 3-velocity of our FIDOs 
relative to the normals on
$\Sigma_t$ is equal to the shift 3-vector 
$\beta^{i}$ in 
   Eq.~(\ref{line.element}).


   {\it The operational definitions of $\vec{E}_{g}$ 
and $\vec{B}_{g}.$---} 
These definitions are independent of perturbation theory.
They involve FIDOs (of any given choice)   
measuring the first time-derivatives along their world lines, 
on the one hand of the momentum components $p_{\hat{i}}$ 
of free-falling quasistatic test particles, 
and on the other hand of the spin components $S_{\hat{i}}$ of gyroscopes 
carried along by the FIDOs,
\begin{eqnarray}
\frac{d}{dt} p_{\hat{i}} \equiv
m E_{\hat{i}}^{g} \quad \mbox{free-falling quasistatic 
test particle,}
\label{op.def.E}
\\
\frac{d}{dt} S_{\hat{i}} \equiv
- \frac{1}{2} [\vec{B}_{g} \wedge \vec{S}]_{\hat{i}}
\quad \mbox{gyro comoving with FIDO,}
\label{op.def.B} 
\end{eqnarray}
where $t$ is the local time measured by the FIDO.
Arrows denote 3-vectors in the tangent spaces spanned by 
the spatial legs of our LONBs.
$\vec{E}_{g} \equiv \vec{g}$ is the gravitational acceleration
of free-falling quasistatic test particles relative to the FIDO.
Eqs.~
   (\ref{op.def.E},\ref{op.def.B}) 
are the same as for a classical charged spinning test particle 
in an electromagnetic field 
except that $q$ is replaced by $m,$
and the gyromagnetic ratio $q/(2m)$ is replaced by $1/2.$
Eq.~
   (\ref{op.def.B}) 
gives the angular velocity of precession of the 
gyroscope's spin axis relative to the axes of the FIDO,
\begin{equation}
\vec{\Omega}_{\rm gyro} \equiv -\frac{1}{2} \vec{B}_{g},
\label{var.op.def.B}
\end{equation}
which is an equivalent operational definition of 
$\vec{B}_{g}.$


\section{Connection Coefficients and the \protect\\          
  equations of motion for matter \protect\\  in gravitomagnetism}
 
  {\it Connection 1-forms.---}
The connection 1-forms 
$ \tilde{\omega}^{\hat{a}}_{\ \hat{b}}$ 
resp their components in LONBs (= Ricci rotation coefficients),
$(\omega^{\hat{a}}_{\ \hat{b}})_{\hat{c}},$  
are defined by
\begin{equation}
\nabla_{\hat{a}} \bar{e}_{\hat{b}} = \bar{e}_{\hat{c}}
(\omega^{\hat{c}}_{\ \hat{b}})_{\hat{a}}.
\end {equation}
In words: the Ricci rotation coefficients 
$(\omega^{\hat{c}}_{\ \hat{b}})_{\hat{a}}$
give the rotation resp the Lorentz boost $(\omega^{\hat{c}}_{\ \hat{b}})$
of the LONBs relative to parallel transport along $\bar{e}_{\hat{a}}.$
Parallel transport is given by 
free fall for $\bar{u}_{\rm FIDO} = \bar{e}_{\hat{0}}$
and gyroscope axes for $\bar{e}_{\hat{i}}.$
Relative to FIDOs the equations of motion for free-falling test particles  
(geodesic equation) and for spin axes of gyroscopes
(Fermi transport) specialized to gyroscopes 
carried along by a FIDO are
\begin{equation}
\frac{dp^{\hat{a}}}{dt} +
(\omega^{\hat{a}}_{\ \hat{b}})_{\hat{c}}
p^{\hat{b}} \frac{dx^{\hat{c}}}{dt}=0, \quad
\frac{dS^{\hat{i}}}{dt} + 
(\omega^{\hat{i}}_{\ \hat{j}})_{\hat{0}} S^{\hat{j}}=0.
\label{geodesic.equation.and.Fermi.transport}
\end{equation}
With Eqs.~(\ref{geodesic.equation.and.Fermi.transport})
the operational definitions 
Eqs.~
   (\ref{op.def.E}, \ref{var.op.def.B})
get translated into the equivalent definitions 
involving connection coefficients with a 
displacement index $\hat{0}$,
namely a Lorentz boost $\omega_{\hat{i} \hat{0}}$ 
per unit time (acceleration $\vec{g}=\vec{E}_{g}$) resp 
a rotation angle $\omega_{\hat{i} \hat{j}}$
per unit time (angular velocity $\vec{\Omega}_{gyro}=
-\frac{1}{2}\vec{B}_{g}$),
\begin{equation}
(\omega_{\hat{i}\hat{0}})_{\hat{0}} \equiv
- E_{\hat{i}}^{g}, \quad 
(\omega_{\hat{i}\hat{j}})_{\hat{0}} \equiv
- \frac{1}{2}  B_{\hat{i}\hat{j}}^{g},
\label{2nd.op.def.EB}
\end{equation}
where $B_{\hat{i}\hat{j}} \equiv 
\varepsilon_{\hat{i}\hat{j}\hat{k}}B_{\hat{k}}.$


 {\it The computation of connection coefficients
in Cartan's formalism.--- The first step} is to express 
our choice of LONBs $\bar{e}_{\hat{a}}(P)$ 
in terms of the coordinate bases 
$\bar{e}_{\alpha}(P)=\partial / \partial x^{\alpha}  ,$ i.e. 
$\bar{e}_{\hat{a}}=(e_{\hat{a}})^{\alpha}\bar{e}_{\alpha}.$
To first order in 
$(\beta^{i}/c)$
\begin{equation}
\bar{e}_{\hat{0}}=\bar{e}_{0}, 
\qquad \bar{e}_{\hat{k}}=
\frac{1}{ah_{k}}(\bar{e}_{k}+\beta_{k}\bar{e}_{0}).
\label{expansion of basis vectors}
\end{equation}
For FRW with $k=0$ in Cartesian spatial coordinates 
all $h_{i}=1$. For FRW in spherical coordinates 
and $k=0, \pm 1$ we have 
$h_{\chi}=1,  \, h_{\theta}= R(\chi), 
h_{\phi}  = R(\chi) \sin \theta$ with 
$R(\chi) \equiv  \{ \chi,  \sin \chi,  \sinh \chi \}.$
Since the spatial LONBs point in the 
same spatial directions as the spatial
coordinate bases, we use latin letters 
from the middle of the alphabet both
for spatial LONBs (with hat) and for 
spatial coordinate bases (without hat).
The dual bases (basis 1-forms) 
$\tilde{\theta}^{\hat{a}}$ for LONBs 
are defined by 
$\langle \tilde{\theta}^{\hat{a}}, \bar{e}_{\hat{b}} \rangle =
\delta^{\hat{a}}_{\hat{b}},$ 
where tildes designate space-time 1-forms.
The LONB 1-forms $\tilde{\theta}^{\hat{a}}$ are expanded in the  
coordinate basis 1-forms, $\tilde{\theta}^{\alpha} = 
\tilde{d}x^{\alpha}$, i.e.                       
$\tilde{\theta}^{\hat{a}}= 
(\theta^{\hat{a}})_{\alpha}\tilde{\theta}^{\alpha},$ by
\begin{equation}
\tilde{\theta}^{\hat{0}}=
\tilde{\theta}^{0}-\beta_{k} \tilde{\theta}^{k},\qquad 
\tilde{\theta}^{\hat{k}}=ah_{k}  \tilde{\theta}^{k}.
\label{expansion of basis 1-forms}
\end{equation} 
The coefficients of the inverse expansion, 
i.e. coordinate bases 
in terms of LONBs, are
$(e_{\alpha})^{\hat{a}} = (\theta^{\hat{a}})_{\alpha}$ resp.
$(\theta^{\alpha})_{\hat{a}} = (e_{\hat{a}})^{\alpha}.$


{\it The second step} is computing the exterior derivative $d$ of 
the basis-1-forms, $(d\theta^{\hat{c}})_{\alpha \beta} \equiv
\partial_{\alpha}(\theta^{\hat{c}})_{\beta} -
\partial_{\beta} (\theta^{\hat{c}})_{\alpha},$
where $[\alpha \beta]$ must be in the coordinate basis.
Then one converts to components $[\hat{a} \hat{b}]$ in the LONB,
\begin{equation}
(d\theta^{\hat{c}})_{\hat{a} \hat{b}} \equiv
- C_{\hat{a} \hat{b}}^{\ \ \hat{c}} =
(e_{\hat{a}})^{\alpha} 
[\partial_{\alpha} (\theta^{\hat{c}})_{\beta}
-\partial_{\beta}  (\theta^{\hat{c}})_{\alpha}]
(e_{\hat{b}})^{\beta}.
\label{exterior.derivative.in.LONB}
\end{equation} 
The coefficients $C_{\hat{a} \hat{b}}^{\ \ \hat{c}}$
are identical to the commutation coefficients of the
basis vectors, $[\bar{e}_{\hat{a}},\bar{e}_{\hat{b}}] \equiv
C_{\hat{a}\hat{b}}^{\ \ \hat{c}} \bar{e}_{\hat{c}}.$
This is easily shown by using 
$\partial_{\alpha} \langle \tilde{\theta}^{\hat{c}}, 
\bar{e}_{\hat{b}} \rangle = 0
= [\partial_{\alpha} 
(\theta^{\hat{c}})_{\beta}] (e_{\hat{b}})^{\beta} +    
(\theta^{\hat{c}})_{\beta} [\partial_{\alpha}(e_{\hat{b}})^{\beta}].$


{\it The third step} is obtaining the connection coefficients 
from the commutation coefficients.
The definition of the connection via basis 1-forms is
$(\nabla_{\alpha} \theta^{\hat{c}})_{\beta}
=-(\omega^{\hat{c}}_{\ \hat{d}})_{\alpha}
(\theta^{\hat{d}})_{\beta}.$
We take both the displacement index $\alpha$ and the equation's 
component index $\beta$ in the coordinate basis,
and we antisymmetrize in $[\alpha \beta].$ This makes 
the Christoffel symbols $\Gamma^{\rho}_{\ \alpha \beta}$ 
on the left-hand side disappear,
since they are symmetric in $\alpha,\beta.$
Hence the left-hand side reduces to 
$(d \theta ^{\hat{c}})_{\alpha \beta}$.
Dropping the equation's component indices 
$ [\alpha,\beta]$ gives
\begin{equation}
\tilde{d} \tilde{\theta}^{\hat{c}}
= - \tilde{\omega}^{\hat{c}}_{\ \hat{d}}
\wedge \tilde{\theta}^{\hat{d}}.
\label{first.Cartan}
\end{equation}
This is Cartan's first equation.
The wedge product (exterior product) of two 1-forms 
$\tilde{\sigma}$ and $\tilde{\rho}$ is
$(\sigma \wedge \rho)_{\alpha\beta}
\equiv 
\sigma_{\alpha}\rho_{\beta}-\sigma_{\beta}\rho_{\alpha}.$
Taking Cartan's first equation in LONB components
gives the commutation coefficients, Eq.~
   (\ref{exterior.derivative.in.LONB}), 
on the left-hand side,
and the right-hand side simplifies in LONB 
because $(\theta^{\hat{d}})_{\hat{b}}=
\delta^{\hat{d}}_{\hat{b}}.$ 
Hence Cartan's first equation in LONB components is
\begin{equation}
C_{\hat{a}\hat{b}}^{\ \ \hat{c}}=
(\omega^{\hat{c}}_{\ \hat{b}})_{\hat{a}} -
(\omega^{\hat{c}}_{\ \hat{a}})_{\hat{b}}.  
\end{equation}
This equation is easily solved for the rotation coefficients,
\begin{equation}
(\omega_{\hat{c}\hat{b}})_{\hat{a}}= \frac{1}{2}[
C_{\hat{c}\hat{b}\hat{a}} +
C_{\hat{c}\hat{a}\hat{b}} -
C_{\hat{b}\hat{a}\hat{c}}].
\label{connection.coeff.commutation.coeff} 
\end{equation}


   {\it Connection coefficients for vorticity perturbations on a Minkowski 
background.---}  
To first order in $\beta^{i}$
and with Cartesian spatial coordinates on $\Sigma_{t},$
the commutation coefficients 
are very simple to compute,
because only $(\theta^{\hat{0}})_{i}=-\beta_{i}$ 
is space-time dependent, and the prefactors in 
   Eq.~(\ref{exterior.derivative.in.LONB}), 
$(e_{\hat{a}})^{\alpha}$ and $(e_{\hat{b}})^{\beta},$
can be set to $1$. 
Hence $C_{\hat{a}\hat{b}}^{\ \ \hat{0}}=
(d\beta)_{ab},$ and
\begin{eqnarray}
    (\omega_{\hat{i}\hat{0}})_{\hat{0}} 
&\equiv& \, - \quad \, E_{\hat{i}}^{g} \quad \, 
  = \quad  \, \, \, \partial_{t}\beta_{i}, \nonumber \\
    (\omega_{\hat{i}\hat{j}})_{\hat{0}} 
&\equiv& -\frac{1}{2}\varepsilon_{\hat{i}\hat{j}\hat{k}}B_{\hat{k}}^{g}
  = - \frac{1}{2} (d \beta)_{ij},
  = (\omega_{\hat{i}\hat{0}})_{\hat{j}}, \nonumber \\
    (\omega_{\hat{i}\hat{j}})_{\hat{k}} 
&=& 0.
\end{eqnarray} 
All components of connection coefficients 
with respect to LONB's are directly measurable 
(in contrast to Christoffel symbols, which refer to
coordinate bases).


   {\it Connection coefficients for vorticity perturbations  
on FRW with} $k=0, \pm 1.$---
It is again straightforward to compute 
the commutation coefficients and the connection coefficients
using Eqs.~
   (\ref{expansion of basis vectors},
    \ref{expansion of basis 1-forms},
    \ref{exterior.derivative.in.LONB}, 
    \ref{connection.coeff.commutation.coeff}),
\begin{eqnarray}
(\omega_{\hat{i}\hat{0}})_{\hat{0}}&\equiv&
- E_{\hat{i}}^{g}=
\frac{1}{ah_{i}} \partial_{t}\beta_{i},
\label{dbeta/dt}
\\
(\omega_{\hat{i}\hat{j}})_{\hat{0}} &\equiv&
-\frac{1}{2}\varepsilon_{\hat{i}\hat{j}\hat{k}}B_{\hat{k}}^{g}=
-\frac{1}{2a^{2}h_{i}h_{j}}(d\beta)_{ij},
\label{curl beta}
\\
(\omega_{\hat{i}\hat{0}})_{\hat{j}} &=&
-\frac{1}{2}\varepsilon_{\hat{i}\hat{j}\hat{k}}B_{\hat{k}}^{g}
+\delta_{\hat{i}\hat{j}}H,
\label{omega.i0j}
\\
(\omega_{\hat{i}\hat{j}})_{\hat{k}}
&=& 
\frac{\delta_{\hat{i}\hat{k}}}{ah_{j}}(H\beta_{j}+
\partial_{j}L_{\hat{i}})-
\frac{\delta_{\hat{j}\hat{k}}}{ah_{i}}(H\beta_{i}+
\partial_{i}L_{\hat{j}}), \quad
\label{omega.ijk}
\end{eqnarray}
where $L_{\hat{i}} \equiv \log h_{i}.$
Since we work to first order in the vorticity 
perturbations (i.e. in $\beta ^{i}$), we can identify 
$\vec{E}_{g}$ and $\vec{B}_{g}$ 
with vectors in $\Sigma_{t}.$        
From Eqs.~(\ref{dbeta/dt}, \ref{curl beta}), 
we see that the shift vector $\vec{\beta}$
must be identified with the gravitomagnetic vector potential
$\vec{A}_{g}$.  
From Eqs.~
   (\ref{dbeta/dt}, \ref{curl beta}) 
follow
\begin{eqnarray}
&&\vec{B}_{g}={\rm curl} \vec{A}_{g}, \quad 
\vec{E}_{g}=-\frac{1}{a}\partial_{t}(a\vec{A}_{g}), \quad
\label{fields.from.vector.potential}
\\
&& \quad \quad {\rm curl} \vec{E}_{g}+
\frac{1}{a^{2}}\partial_{t}(a^{2}\vec{B}_{g})=0.
\label{Faraday}
\end{eqnarray}
These equations are identical with the hogeneous equations for 
electromagetism in FRW space-times with $k=0, \pm 1$.


   {\it Equation of motion for free-falling test particles.---}  
The equation of motion (geodesic equation) for test particles of
arbitrary velocities $v \leq c$  
in linear vorticity perturbations on a {\it Minkowski background} reads
\begin{equation}
\frac{d}{dt}(p_{\hat{i}}) =
\varepsilon [\vec{E}_{g}+
(\vec{v} \wedge \vec{B}_{g})]_{\hat{i}},
\label{Lorentz.law}
\end{equation}
identical with the one for electromagnetism, 
except that the charge $q$ 
is replaced by the energy $\varepsilon$ of the test particle.
With Eq.~(\ref{var.op.def.B}) 
and in a stationary gravitomagnetic field ($\vec{E}_{g}=0$) 
Eq.~(\ref{Lorentz.law}) becomes 
$\frac{d}{dt}(p_{\hat{i}}) = 
- 2 \varepsilon [\vec{v} \wedge \vec{\Omega}_{\rm gyro}]_{\hat{i}},$
the Coriolis force law. Note that $\Omega_{\rm gyro}$ is minus the
rotation velocity of the FIDO relative
to the gyroscopes' axes.
A homogeneous gravitomagnetic field 
can be transformed away completely by going to a rigidly rotating
coordinate system, i.e. physics in a homogeneous gravitomagnetic field
is equivalent to physics on a merry-go-round in Minkowski space.---
Note that there are no terms bilinear in $\vec{v}$ 
for test particles of arbitrary velocities $v \leq c.$


For {\it FRW with $k=0$} we obtain
\begin{equation}
\frac{1}{a} \frac{d}{dt} (a p_{\hat{i}})=
\varepsilon [\vec{E}_{g}+
(\vec{v} \wedge \vec{B}_{g})    
+ H \vec{v} \wedge (\vec{\beta} \wedge \vec{v})]_{\hat{i}}.
\end{equation}
For FRW with $k \pm 1$ 
there are additional terms from
$\partial_{\hat{i}}L_{\hat{j}}$ in Eq.~
   (\ref{omega.ijk}).
These terms are present even in the absence of
vorticity perturbations and of Hubble expansion, 
because in the spherical basis spatial LONBs are not
parallelized.


\section{Einstein's Equations for Gravitomagnetism: Amp\`ere's Law}

   {\it Curvature.---}The curvature 2-form 
${\tilde{\cal R}}^{\hat{a}}_{\ \hat{b}}$
has LONB components 
$({\cal R}^{\hat{a}}_{\ \hat{b}})_{\hat{c}\hat{d}},$
which are the LONB components of the 
Riemann tensor $R^{\hat{a}}_{\ \hat{b}\hat{c}\hat{d}}.$
The Riemann tensor can be operationally defined 
by the action of 
$(\nabla_{\gamma} \nabla_{\delta}
- \nabla_{\delta} \nabla_{\gamma})$
on the LONB 1-form
$\tilde{\theta}^{\hat{a}},$
\begin{equation}
(\nabla_{\gamma} \nabla_{\delta}
- \nabla_{\delta} \nabla_{\gamma})
\tilde{\theta}^{\hat{a}} =
- \tilde{\theta}^{\hat{b}} 
({\cal R}^{\hat{a}}_{\ \hat{b}})_{\gamma\delta},
\end{equation}
where the covariant derivatives $\nabla_{\gamma}$
and $\nabla_{\delta}$
must be in the coordinate basis.
To compute the curvature 2-form we first use
$\nabla_{\delta} \tilde{\theta}^{\hat{a}}=
-(\omega^{\hat{a}}_{\ \hat{e}})_{\delta} 
\tilde{\theta}^{\hat{e}}.$
Then we let this right-hand side be acted on by 
$\nabla_{\gamma}.$ This gives two terms.
One term comes from $\nabla_{\gamma}$ acting on 
$\tilde{\theta}^{\hat{e}},$ 
and after antisymmetrization in $[\gamma \delta]$
it produces 
$-(\omega^{\hat{a}}_{\ \hat{e}} \wedge 
\omega^{\hat{e}}_{\ \hat{b}})_{\gamma\delta} 
\tilde{\theta}^{\hat{b}}.$
The other term comes from $\nabla_{\gamma}$ acting on
the expansion coefficient (number field) 
$(\omega^{\hat{a}}_{\ \hat{e}})_{\delta},$ 
where it can be replaced by $\partial_{\gamma},$
and after antisymmetrization in $[\gamma \delta]$
it produces 
$-(d \omega^{\hat{a}}_{\ \hat{b}})_{\gamma \delta}
\tilde{\theta}^{\hat{b}}.$
Hence we obtain
\begin{equation}
\tilde{{\cal R}}^{\hat{a}}_{\ \hat{b}} =
\tilde{d} \tilde{\omega}^{\hat{a}}_{\ \hat{b}} +
\tilde{\omega}^{\hat{a}}_{\ \hat{e}} \wedge 
\tilde{\omega}^{\hat{e}}_{\ \hat{b}},
\end{equation}
which is Cartan's second equation.


   {\it Cartan's 2nd equation in LONB.---}
To obtain the LONB components of 
the first term of the right-hand side,
$(d\omega^{\hat{a}}_{\ \hat{b}})_{\hat{c}\hat{d}},$
we must first convert the connection components of 
Eqs.~
   (\ref{dbeta/dt} - \ref{omega.ijk}) 
from the LONB to the coordinate basis,
$(\omega^{\hat{a}}_{\ \hat{b}})_{\delta}=
 (\omega^{\hat{a}}_{\ \hat{b}})_{\hat{d}}
(\theta^{\hat{d}})_{\delta},$   
then take the exterior derivative,
$\partial_{\gamma}\{(\omega^{\hat{a}}_{\ \hat{b}})_{\hat{d}}
(\theta^{\hat{d}})_{\delta}\} - [\gamma \leftrightarrow \delta],$   
and then convert the result 
back from coordinate components to LONB components.
The partial derivative of the product gives two terms,
one with a partial derivative of 
$(\omega^{\hat{a}}_{\ \hat{b}})_{\hat{d}},$
the other with $\partial_{\gamma} (\theta^{\hat{d}})_{\delta},$
which produces another connection 1-form component.
In the second term of Cartan's second equation these conversions 
from LONB to coordinate basis and back again cancel, 
since there is no derivative in between.
The result is Cartan's 2nd equation in LONB components,
\begin{eqnarray}
({\cal R}^{\hat{a}}_{\ \hat{b}})_{\hat{c}\hat{d}} &=&
[(e_{\hat{c}})^{\gamma}
\partial_{\gamma} (\omega^{\hat{a}}_{\ \hat{b}})_{\hat{d}} 
- (\omega^{\hat{a}}_{\ \hat{b}})_{\hat{f}} 
(\omega^{\hat{f}}_{\ \hat{d}})_{\hat{c}}
\nonumber 
 + (\omega^{\hat{a}}_{\ \hat{e}})_{\hat{c}} 
  (\omega^{\hat{e}}_{\ \hat{b}})_{\hat{d}}] \\
&&  - [\hat{c} \leftrightarrow \hat{d}].
\label{curvature.computation}
\end{eqnarray}


   {\it Einstein equations for vorticity perturbations 
of Minkowski space.---}
For linear vorticity perturbations of Minkowski space
(with Cartesian coordinates for 3-space)
all non-zero connection coefficients in
Eqs.~(\ref{dbeta/dt} - \ref{omega.ijk}) are of first order
in the perturbations. 
Therefore the second term of Cartan's second equation
can be neglected, and in the first term 
one need not distinguish 
components in LONB from components in the coordinate basis. 
For vorticity perturbations the important Einstein equation 
is the equation for
$G_{\hat{0}\hat{i}}=R_{\hat{0}\hat{i}},$
\begin{equation} 
R_{\hat{0}\hat{i}}=
({\cal R}_{\hat{0}\hat{j}})_{\hat{i}\hat{j}}=
(d\omega_{\hat{0}\hat{j}})_{\hat{i}\hat{j}}=
\frac{1}{2}({\rm curl} \vec{B})_{\hat{i}}.
\end{equation}
Hence Einstein's $G_{\hat{0}\hat{i}}$-equation 
for vorticity perturbations in Minkowski space reads
\begin{equation}
{\rm curl} \vec{B_{g}} = -16 \pi G_{N} \vec{J}_{\varepsilon}.
\label{Ampere}
\end{equation}
$J_{\varepsilon}^{\hat{i}} \equiv T^{\hat{0}\hat{i}} = 
(\rho + p) v^{\hat{i}}$ 
is the energy current density, which is equal 
to the momentum density. 
Eq.~(\ref{Ampere}) is identical to the 
original law of Amp\`ere for magnetism, 
except that the charge current $\vec{J}_{q}$ 
is replaced by the energy current $\vec{J}_{\varepsilon},$ 
and the prefactor $4 \pi$ is 
replaced by the prefactor $(-16 \pi G_{N}).$
In contrast to the Amp\`{e}re-Maxwell equation, the Maxwell term
$(\partial_{t} \vec{E})$ is absent in gravitomagnetodynamics. 
The $G_{\hat{i}\hat{0}}$ is an equation at fixed time, 
a constraint equation, called momentum constraint, 
since the momentum density appears on the
right-hand side of Eq.~
   (\ref{Ampere}).

To see the analogous structures of gravitomagnetism 
and electromagnetism, it is more instructive to formulate
this constraint equation, as we have done in Eq.~(\ref{Ampere}), 
via the connection 1-forms, which involves 
$(\partial_{i} \beta_{j} - \partial_{j} \beta_{i}),$
i.e. the gravitomagnetic field,
than via the extrinsic curvature tensor $K_{ij},$
which involves
$(\partial_{i} \beta_{j} + \partial_{j} \beta_{i}).$
Of course the resulting constraint, if written in terms of
$\vec{A}_{g}=\vec{\beta},$ is the same, 
$\Delta \vec{A}_{g}= 16 \pi G_{N} \vec{J}_{\varepsilon}.$

   The $G_{ \hat{0} \hat{0} }$ equation with
the source $T_{ \hat{0} \hat{0} }$ is trivially fulfilled,
since these objects are 3-scalars and therefore 
vanish in the vector sector.---
The source $T_{ \hat{i} \hat{j} }$ vanishes, 
since it is of second order in the perturbation.
The $G_{ \hat{i} \hat{j} }$ equations give 
$\partial_{0} (\partial_{i} \beta_{j} + \partial_{j} \beta_{i})=0,$
i.e. the shear of the field 
$\vec{\beta}$ has vanishing time-derivative.


   {\it Einstein equations for vorticity perturbations of spatially flat 
FRW space.---} 
With Cartesian comoving coordinates for flat 3-space
there are two new terms in the connection coefficients,  
$(\omega_{\hat{i}\hat{0}})_{\hat{j}}^{\rm FRW}=
\delta_{\hat{i}\hat{j}}H$ and 
$(\omega_{\hat{i}\hat{j}})_{\hat{k}}
= H( \delta_{\hat{i}\hat{k}}  \beta_{\hat{j}} -
     \delta_{\hat{j}\hat{k}}  \beta_{\hat{i}} ).$
Computing  
$R_{\hat{0}\hat{i}}=
({\cal R}_{\hat{0}\hat{j}})_{\hat{i}\hat{j}}$
with 
   Eq.~(\ref{curvature.computation})
we obtain the corresponding Einstein equation,
\begin{equation}
{\rm curl} \vec{B}_{g} - 4 \dot{H} \vec{A}_{g} 
= - 16 \pi G_{N} \vec{J}_{\varepsilon},
\label{Ampere.FRW}
\end{equation}
where we have used $\vec{\beta}=\vec{A}_{g}.$ 
The scale factor $a$ of the spatially flat FRW universe
does not appear in these equations. 
From $\dot{H}=-4\pi G_N (\rho + p)$ we see that $\dot{H} \leq 0$ for
$p\geq -\rho.$ Therefore we define the $H-$dot radius by
$R_{\dot{H}}^2 = (-\dot{H})^{-1}$, and we define $\mu^2=-4\dot{H} = 
(\frac{1}{2} R_{\dot{H}})^{-2}$. 
We insert the vector
potential $\vec{A}_g=\vec{\beta}$, 
we use ~${\rm div}~\vec{A}_g =0,$ 
hence ~${\rm curl~curl}~\vec{A}_g = - \Delta~\vec{A}_g.$ 
Therefore Eq.~(\ref{Ampere.FRW}) becomes
\begin{equation}
\left(- \Delta \,
+\,\mu^2\right) \vec{A}_g \;=\; - 16\,\pi\; G_N
\;\vec{J}_\varepsilon~. 
\label{Ampere.FRW.A}
\end{equation}
The new term on the left-hand side, $(- 4
\dot{H}\;\vec{\beta})=(\mu^2 \vec{A}_g)$, dominates for
superhorizon perturbations.


\section{Mach's Principle}

{\it Our first result.---}
The solution of 
     Eq.~(\ref{Ampere.FRW.A})   
is the Yukawa potential for
$\vec{A}_g=\vec{\beta}$
in terms of the sources~ $\vec{J}_\varepsilon$ at the same fixed time,
\begin{equation}
\vec{A}_g (\vec{r},t)=
-4\, G_N \int d^3
r'\,\vec{J}_\varepsilon ( \vec{r'}, t)~ \frac{{\rm
    exp}( - \mu |\vec{r}-\vec{r'}|)}{|\vec{r}-\vec{r'}|}.
\label{Yukawa.potential}
\end{equation}
This is analogous to the formula for ordinary magnetostatics 
except for the exponential cutoff. 
The Green function which is exponentially growing for 
$r' \rightarrow \infty$ is rejected on the standard grounds 
of field theory.
The Yukawa potential in 
    Eq.~(\ref{Yukawa.potential}) 
has an exponential cutoff for
$|\vec{r}-\vec{r'}| \geq  1/ \mu$. This gives our first
important conclusion: The contributions of vorticity perturbations
beyond the $H-$dot radius are exponentially suppressed.

{\it Our second result} concerns the exact dragging of gyroscope axes by a
homogeneous rotation of cosmological matter out to significantly
beyond the $H-$dot radius (for the exponential cutoff to be
effective). This holds for any equation of state. 
This is easily seen
from Einstein's $G_{\hat{0}\hat{\jmath}}$ equation
   (\ref{Ampere.FRW.A})
in $k$-space for superhorizon perturbations, 
$k_{\rm phys}^2 \;\ll\;(-\dot{H})$,
where the $\Delta$-term can be dropped. Using
$\vec{J}_\varepsilon \;=\; (\rho + p)\,\vec{v}_{\rm fluid}$ ~~~ and
$\dot{H} \;=\; - 4 \pi\;G_N (\rho+p)~$
we see that all the prefactors cancel, 
and we obtain $\vec{\beta}(\vec{x}) = -
\vec{v}_{\rm fluid}(\vec{x}).$ 
With $\vec{\Omega}_{\rm gyroscope} = - 
- \frac{1}{2}(\vec{\nabla} \times \vec{\beta})$
and with $\vec{\Omega}_{\rm
  fluid} = \frac{1}{2} ( \vec{\nabla} \times \vec{v}_{\rm fluid})$ 
we obtain the result that for $R_{\rm pert} \gg R_{\dot{H}}$
there is an exponentially fast approach to
\begin{equation}
\vec{\Omega}_{\rm gyroscope} = \vec{\Omega}_{\rm matter}.
\label{exact.dragging}
\end{equation}
This proves exact dragging of gyroscope axes here by a homogeneous
rotation of cosmological matter out to significantly beyond the
$\dot{H}$ radius. 

{\it Our third result} concerns the most general vorticity perturbation in
linear approximation, and it states what {\it specific average} of energy
flow in the universe determines the motion of gyroscope axes here at
$r=0.$ We take  $\vec{B}_{g}={\rm curl} \, \vec{A}_{g}$ of 
   Eq.~(\ref{Yukawa.potential}), 
we set $r=0,$ which gives
$\vec{B}_{g}(r=0) = - \frac{1}{2}\Omega_{\rm gyro},$
and obtain the equation for $\Omega_{\rm gyro}$ in terms of the
sources at the same fixed time,
\begin{eqnarray}
&& \vec{\Omega}_{\rm gyro} \,=\, \nonumber \\
&& 2 G_N (\rho +p) \int d^3 
 r\, \frac{1}{r^3}\, \left[\left( 1+\mu r\right) \,
  e^{-\mu\,r}\right] \left[ \vec{r} \wedge \vec{v} \left( \vec{r}
    \right)\right]. \quad  \quad
\label{Mach}
\end{eqnarray}
{\it This is the precise expression for Mach's principle,}
it says exactly what average of the motion of energy
in the universe determines  $\vec{\Omega}_{\rm gyro}.$
Mach had asked: 
{\it ``What share has every mass in the determination 
of direction in the law of inertia?} 
No definite answer can be given by our experiences''
   \cite{Mach.Mechanik}.---
In the integrand we have the gravitomagnetic moment density
$\vec{\mu}_{g}=\frac{1}{2} (\vec{r} \wedge \vec{J}_{\varepsilon}),$
analogous to
the magnetic moment density of an electric current distribution.
The expression $(\vec{r} \wedge \vec{J}_{\varepsilon})= 
(\vec{L}_{\varepsilon})$ is  
the measured angular momentum.
This is the lowest term, the
$\ell=1 $ term, in the multipole expansion of the source
for $r_{\rm obs} = r_{\rm gyro}=0$ and $
r_{\rm source} > 0.$ Higher multipoles cannot contribute to
the gravitomagnetic field $\vec{B}_{g}$
at $ r = 0.$
At each radius $r$ only a term equivalent 
to a rigid rotation with angular velocity $\vec{\Omega}(r)$
contributes in 
   Eq.~(\ref{Mach}).
Using this we obtain
\begin{equation}
\vec{\Omega}_{\rm gyro} = 
\frac{4}{3} \int \frac{dr \, r}{R^{2}_{\dot{H}}} \, 
\vec{\Omega}_{\rm matter}(r) \, C_{\mu}(r),
\label{Mach.2}
\end{equation}
where $C_{\mu}(r)$ is the cutoff function, 
i.e. the first square bracket in Eq.~(\ref{Mach}).
For the special case $\vec{\Omega}_{\rm matter}(r)$ 
independent of $r$ we recover 
   Eq.~(\ref{exact.dragging}).


   {\it The $r$-dependence of the weight function} 
of the energy current 
$\vec{J}_{\varepsilon}(\vec{r})$ in the integral for 
$\vec{B}_{g}(0)$ for $r \ll R_{\dot{H}}$
is the same $(1/r^{2})$ law as in Amp\'ere's law.
Hence the weight function for the measured angular momentum
$(\vec{r} \wedge \vec{J}_{\varepsilon})$ is  $(1/r^{3}).$
For fixed $\vec{\Omega}$ 
the energy current density increases linearly with $r,$
and the weight function per unit $r$ is the first power of $r,$
   Eq.~(\ref{Mach.2}).
Therefore a perturbation which is 
a rigid rotation out to $R_{\rm pert}$ and zero
outside has a dragging fraction $f_{\rm drag},$ 
which grows quadratically with $R_{\rm pert},$ until
$f_{\rm drag}$ reaches a value near $1$ 
for $R_{\rm pert}= R_{\dot{H}}.$ For $R_{\rm pert}$ increasing
beyond $R_{\dot{H}}$
the dragging fraction approaches the value $1$ 
exponentially fast.


   {\it Dark energy with} $p/ \rho = - 1,$ 
i.e. a cosmological constant,
does not contribute in Mach's principle,
   Eq.~(\ref{Mach}),
since there is no flow of energy 
associated with it, its energy current 
$\vec{J}_{\varepsilon}=(\rho + p) \vec{v}$ vanishes.


   {\it Measuring everything relative to axes of gyroscopes 
at a given location} 
(instead of relative to galaxies in the asymptotic FRW space)
makes the the left-hand side of 
   Eq.~(\ref{Mach}) 
vanish. Hence 
   Eq.~(\ref{Mach}) 
reduces to the statement that the angular momentum of matter,
measured relative to the gyroscopes at the given location, 
will vanish after averaging with
the weight $r^{-3}$ and with the exponential cutoff $C_{\mu}(r).$
From the point of view of measurements
it is preferrable to mesure relative to gyros at one given location.
But to see the structure of gravitomagnetism most clearly,
it is best to measure relative to galaxies in the asymptotic FRW space.


   {\it Analogous dragging effects in magnetostatics}
(e.g. a rotating charged spherical shell acting on  
magnetic dipole moments inside), 
have the opposite sign in Amp\`ere's law compared to
gravitomagnetism, Eq.~(\ref{Ampere}),  
and this causes antidragging in the
   magnetostatic case.


   {\it Einstein's objection to Mach's principle in his 
autobiographical notes of 1949.---} 
Einstein wrote 
   \cite{Einstein}: 
"Mach conjectures that inertia would have to depend
upon the interaction of masses, precisely as was true for Newton's
other forces, a conception which for a long time I considered as in
principle the correct one. It presupposes implicitly, however, that
the basic theory should be of the general type of Newton's mechanics :
masses and their interactions as the original concepts.  The attempt
at such a solution does not fit into a consistent field theory, as
will be immediately recognized."  
We have shown, how this apparent difficulty
is resolved in General Relativity, 
specifically in weak Gravitomagnetism:
The relevant Einstein equation has the form of 
Amp\`ere's law with a Yukawa term, 
   Eq.~(\ref{Ampere.FRW}), 
hence the measured mass-energy flow out there in the universe 
does indeed determine the precession of gyroscope axes here, 
   Eq.~(\ref{Mach}).  


\section{Measured Matter Input for \protect\\     
         Einstein's Equations and for \protect\\  Mach's Principle}

The {\it input} on the right-hand side of 
   Eq.~(\ref{Mach}),
which expresses Mach's principle, is the {\it measured} 
angular velocity of matter (stars, galaxies, etc), 
measured relative to galaxies in the asymptotic FRW space 
(``asymptotic galaxies'').
No knowledge of the metric perturbation
$\vec{\beta}$ is needed when determining the input for 
   Eq.~(\ref{Mach}).
This is a purely {\it kinetic input},  
which means that it is directly determined by the 
{\it measured} state of motion of matter at a given time without 
knowing the metric perturbation $\vec{\beta}.$
Similarly the 
{\it measured angular momentum} is a purely kinetic 
input,
$(\rho + p) (\vec{r} \wedge \vec{v}) r^{2}dr = 
 (\rho + p) r^{2} \vec{\Omega}       r^{2}dr .$ 
The {\it dynamical output} of solving Einstein's $G_{\hat{0}\hat{i}}$ 
equation is  
$\vec{\beta}= \vec{A}_{g},$ the gravitomagnetic vector potential, in
    Eq.~(\ref{Ampere.FRW.A}) 
and the gravitomagnetic force $\vec{B}_{g}$ in
    Eq.~(\ref{Mach}).
The measured, {\it kinetic angular momentum} must be distinguished from
the {\it canonical angular momentum}, which we introduce (review) 
in the following paragraphs.


     {\it Action and Lagrangian for Gravitomagnetism.---}
The Einstein-Hilbert action 
for linear vorticity perturbations 
on a Minkowski background and for point particles is
\begin{eqnarray}
S  &=&  (16 \pi G_{N})^{-1} \int d^{4}x \, \sqrt{g} \, \, 
           ({\rm curl} \, \vec{A}_{g})^{2} \nonumber \\
   &+& \int dt \sum_{n} \, [\frac{1}{2} m \, \dot{\vec{x}}_{n}^{2} 
        + m \, \dot{\vec{x}}_{n} \, \vec{A}_{g}(\vec{x}_{n},t)] \quad
\label{action}
\end{eqnarray}
Except for the prefactor $(16 \pi G_{N})^{-1}$ and the sign 
of the first term, the action is the same as for electromagnetism without
the $(\partial_{t} \vec{A})^{2}$-term.---
Generally the equations of motion are given by the Lagrangian via 
the standard Euler-Lagrange equations (as in  
classical mechanics and classical electrodynamics),
if and only if the Lagrangian is defined by 
\begin{equation}
S = \int dt L
\end{equation}
without any metric factors in the integrand.
Hence the Lagrangian for a point particle 
in a gravitomagnetic field is given by the
square bracket of the matter term in Eq.~(\ref{action}).


{\it The canonical momentum} is defined by 
$(p_{\rm can})_{k} = \partial L / \partial \dot{x}^{k},$ where $k = 1, 2, 3.$
From Eq.~(\ref{action}) we obtain in Cartesian coordinates 
\begin{equation}
\vec{p}_{\rm can} = m (\dot{\vec{x}} + \vec{A}_{g}).
\label{can.mom}
\end{equation}
This is the same equation as in classical mechanics 
for point particles in an electromagnetic field,
except that the electric charge $q$ is replaced by $m$.---
The {\it kinetic momentum} is $m\dot{\vec{x}}.$ 
It is directly measured,
it can be used as the {\it input} for solving Einstein's
equations, because it is
independent of the gravitomagnetic vector potential 
$\vec{A}_{g} =  \vec{\beta},$ which is an output
of solving Einstein's equations. 
On the other hand the {\it canonical momentum} (for a given 
measured state of motion) depends on the 
gravitomagnetic vector potential $\vec{A}_{g}.$
Therefore the canonical momentum {\it cannot} be used 
as a {\it input} for solving Einstein's equations.
{\it The canonical momentum cannot be determined   
by a FIDO from measurements before having solved Einstein's equations.}---
In curvilinear coordinates of 3-space 
the general definition of the canonical momentum, 
$(p_{\rm can})_{k} = \partial L / \partial \dot{x}^{k},$ gives
\begin{equation}
(p_{\rm can})_{k} = 
m \, g_{k n} \, (\dot{x}^{n} + A_{g}^{n}) \quad \quad k,n = 1, 2, 3.
\label{can.mom.curvilin}
\end{equation}
From its definition via the Lagrangian,   
the canonical momentum is a 1-form in 3-space, 
i.e. it has a lower 3-index.---  
On the other hand we can also start from the 4-velocity $u^{\nu},$ 
which is the archetype of a 4-vector,  
multiply with the mass to obtain the 4-momentum $p^{\nu},$ 
and pull down the 4-index
with $^{(4)}g_{\kappa \nu }.$ 
This gives $p_{k} = g_{k \nu} (m u^{\nu})= 
m \, g_{k n} (\dot{x}^{n} + A_{g}^{n}),$ which is the same as
the canonical momentum in Eq.~(\ref{can.mom.curvilin}).---
Going to spherical coordinates we note that
$p_{\phi}$ has the physical meaning of
canonical angular momentum.


      {\it In the special case of azimuthal symmetry}, i.e. when 
$\bar{e}_{\phi}=\bar{\xi}$ 
is a rotational Killing vector, 
the canonical angular momentum 
$ p_{\phi}=\langle \tilde{p}, \bar{\xi} \rangle $
is conserved, 
while the measured, kinetic angular momentum of matter,
$(rp_{\hat{\phi}})$,
is not conserved, because of the gravitoelectric induction field
$\vec{E}_{g},$ 
     Eq.~(\ref{Faraday}),
which acts in the $\phi$-direction.
The canonical angular momentum is
relevant for the time-evolution, i.e. for the {\it dynamics,}
and conservation laws, 
not for kinematics (measurements at a given time).


     {\it In the continuum description of matter} 
(energy current density)
the LONB components $T_{\hat{0} \hat{k}}$ can be used as {\it input}
for solving Einstein's equations, since they can be measured without
knowing the output $\beta_{\hat{k}}$ of Einstein's equations.
On the other hand the coordinate-basis components
$T^{0}_{\ k}$ cannot be used as an input, because they cannot be
determined by measurements without a knowledge of $\beta_{k}.$


     {\it In Mach's principle} the input is the 
observations of the angular velocities of stars and galaxies,
and from there the measured {\it kinetic} angular momentum
$(\rho + p)[\vec{r} \wedge \vec{v}]$ as shown in Eq.~(\ref{Mach}).---
Bi\v{c}\'ak  et al 
   \cite{Bicak} 
have proposed to use $p_{\phi},$ the {\it canonical} angular momentum
as the input
on the right-hand side of the Einstein equation.
For the reasons given above in this section, 
we consider this to be a fundamental mistake.
If one does this, the $\dot{H}\vec{A}_{g}-$term on the 
left-hand side of the Einstein equation (\ref{Ampere.FRW}) 
gets cancelled, and the exponential suppression factor disappears.


   {\it Einstein's objection to Mach's Principle 
in his letter to Felix Pirani of 2 February 1954}  
     \cite{Einstein.Pirani}, 
which is quoted by Ehlers in
    \cite{Barbour.Pfister.page93}: 
``If you have a tensor $T_{\mu\nu}$ and not a metric,
then this does not meaningfully describe matter.
There is no theory of physics so far, which can describe
matter without already the metric as an ingredient of the 
description of matter. Therefore within existing theories
the statement that the matter by itself determines the metric
is neither wrong nor false, but it is meaningless.''
We agree with this statement, as long as the components 
of $T_{\mu \nu}$ are given in a coordinate basis,
as e.g. $T^{0}_{\ \phi}$ in the proposal of  Bi\v{c}\'ak  et al
   \cite{Bicak}.
But we disagree with Einstein's objection, 
if the components are given in a LONB,
$T_{\hat{a}\hat{b}},$ because these components 
can be directly measured by the FIDOs.
The FIDO only needs
clocks, meter sticks, and markers, which give 
the directions of his spatial axes. The FIDO has only 
the metric of Special Relativity, 
$\eta_{\hat{m}\hat{n}}= diag \{ -1,+1,+1,+1 \},$
available in the tangent space at his space-time point.
But the FIDO does not need any more structure, not a connection and
not the metric potential functions $g_{\mu \nu} (x),$
which in our case are given by the gravitomagnetic 
vector potential $\vec{A}_{g},$
the solution and output of Einstein's equations. 
Therefore we disagree with Einstein's objection quoted above
(from which he concluded that one should no longer 
speak of Mach's Principle at all), 
if the components of the energy-momentum tensor are given in a LONB.


\section{The local vorticity measured by non-rotating observers}

This quantity is defined as ${\rm curl} \, \vec{v}_{\rm fluid}$ measured
in the local inertial coordinate system which is
comoving with the fluid, i.e. measured relative to the axes of
local gyrosopes (``local compass of inertia''). 
In general coordinates the local vorticity measured 
by non-rotating observers is
\begin{equation}
\omega^{\alpha}= - \varepsilon^{\alpha\beta\gamma\delta}
u_{\beta} \nabla_{\gamma} u_{\delta},
\end{equation} 
where $\varepsilon^{\hat{0}\hat{1}\hat{2}\hat{3}} \equiv -1.$
Mach's principle, as formulated as a general hypothesis by Mach 
and made precise in 
   Eq.~(\ref{Mach}), 
states that the gyroscope axes here follow the rotational flow 
$\vec{J}_{\varepsilon}$ of 
{\it matter in the universe averaged} with a $r^{-2}$ weight and 
an exponential cutoff at the $\dot{H}$ radius.
In general the gyroscope axes here most definitely do not follow the
motion of the {\it local fluid here}, i.e. relative to gyroscopes
the {\it local vorticity is nonzero according to Mach's principle} 
in general.
There is a special case, a rigid rotation of a fluid 
out to a perturbation radius $R_{\rm pert}.$ 
In the limit $R_{\rm pert}/R_{\dot{H}} \rightarrow \infty$
the local vorticity measured by non-rotating observers 
vanishes. But this limit is uninteresting, 
since it produces an unperturbed FRW universe.---
Unfortunately many authors have considered the vanishing of the
vorticity relative to the local compass of inertia to be
a test for Mach's principle.
See e.g. Ozsv\`ath and Sch\"ucking's solution and
discussion of a Bianchi IX model for perturbations of 
the original closed and static Einstein universe
   \cite{Ozsvath}.
In contrast we conclude that the vanishing of the vorticity 
relative to the local compass of inertia is 
not relevant as a test of Mach's principle. 
\bibliography{basename of .bib file}

\end{document}